\documentclass[%
 reprint,
superscriptaddress,
showpacs,
 amsmath,amssymb,
 aps,
floatfix,
]{revtex4-1}
\usepackage{graphicx,xcolor,wrapfig}
\usepackage{dcolumn}
\usepackage{bm}

\usepackage{ulem} 

\begin{document}


\title{Antiphase Oscillations in the Time-Resolved Spin Structure Factor of a Photoexcited Mott Insulator}

\author{Kenji Tsutsui}
\affiliation{Synchrotron Radiation Research Center, National Institutes for Quantum and Radiological Science and Technology, Hyogo 679-5148, Japan}

\author{Kazuya Shinjo}
\affiliation{Department of Applied Physics, Tokyo University of Science, Tokyo 125-8585, Japan}

\author{Takami Tohyama}
\affiliation{Department of Applied Physics, Tokyo University of Science, Tokyo 125-8585, Japan}

\date{received 12 October 2020; revised 14 December 2020; accepted 12 February 2021}
             

\begin{abstract}
Motivated by the recent development of time-resolved resonant-inelastic x-ray scattering (TRRIXS) in photoexcited antiferromagnetic Mott insulators, we numerically investigate momentum-dependent transient spin dynamics in a half-filled Hubbard model on a square lattice.  
After turning off a pumping photon pulse, the intensity of a dynamical spin structure factor temporally oscillates with frequencies determined by the energy of two magnons in the antiferromagnetic Mott insulator.  
We find an antiphase behavior in the oscillations between two orthogonal momentum directions, parallel and perpendicular to the electric field of a pump pulse.  
The phase difference comes from the $B_{1g}$ channel of the two-magnon excitation.  
Observing the antiphase oscillations will be a big challenge for TRRIXS experiments when their time resolution will be improved by more than an order of magnitude.
\end{abstract}
\maketitle



Ultrafast optical pulses are now a common tool for investigating novel nonequilibrium phenomena in strongly correlated electron systems.  
One of the typical methods is a pump-probe optical technique, which has widely been applied to various strongly correlated materials~\cite{Giannetti2016,Nicoletti2016}.  
The pump-probe technique can detect two-particle excitation but cannot see momentum-dependent collective excitation due to negligibly small momentum transfer of probe photons.  
On the other hand, time- and angle-resolved photoemission technique~\cite{Schmitt2008,Rohwer2011,Smallwood2012} can probe momentum-dependent electronic excitation after pumping, but it is limited to a single-particle process where collective excitation appears indirectly. 

The recent development of time-resolved resonant-inelastic x-ray scattering (TRRIXS) opens a new avenue for probing collective two-particle excitation, from which one can investigate novel photoinduced nonequilibrium phenomena in the wide-range of momentum and energy spaces~\cite{Dean2016,Mitrano2019a,Mitrano2019b,Mazzone2020}.  
RIXS can probe not only charge excitation but also magnetic excitation if one uses incident x rays tuned for $L$ edge in transition metals.  
If the lifetime of an intermediate state in the $L$-edge RIXS process is short enough, the dominant contribution to the RIXS spectrum comes from the dynamical charge and spin structure factors~\cite{Brink2006,Brink2007,Ament2007}.  
It is also numerically shown that even for a realistic lifetime scale of an intermediate state in cuprate materials, the magnetic excitation in RIXS gives information on the dynamical spin structure factor~\cite{Tohyama2018}.  
Therefore, TRRIXS is an ideal tool for characterizing transient spin dynamics.  
In fact, TRRIXS has been applied for two-dimensional antiferromagnetic (AFM) Mott insulating materials, Sr$_2$IrO$_4$~\cite{Dean2016} and Sr$_3$Ir$_2$O$_7$~\cite{Mazzone2020}.  
The transient change and recovery of spin-wave excitation have been reported in the timescale of (sub)picosecond.  
These pioneering works open a new window to study novel phenomena in photoexcited AFM Mott insulators. 

In this Letter, we theoretically investigate momentum-dependent spin excitation that evolves after pumping within a femtosecond timescale in the AFM Mott insulator on a square lattice.  
Using a numerically exact-diagonalization technique for a half-filled Hubbard model, we find novel momentum-dependent transient spin dynamics.  
In particular, we demonstrate characteristic temporal oscillations for the intensity of the dynamical spin structure factor, showing an antiphase behavior for two orthogonal directions that are parallel and perpendicular to the electric field of a pump pulse.  
Their oscillation period in time is determined by two-magnon excitation in the Mott insulator.  
This theoretical prediction will be confirmed for Mott insulating cuprates and iridates once TRRIXS is ready for a femtosecond timescale.

In order to describe Mott insulating states in cuprates, we take a single-band Hubbard model on a square lattice at half filling given by
\begin{equation}
H_0=-t_h\sum_{i\delta\sigma} c^\dagger_{i\sigma} c_{i+\delta\sigma}
 -t_h^\prime\sum_{i\delta'\sigma} c^\dagger_{i\sigma} c_{i+\delta'\sigma}
 + U\sum_i n_{i\uparrow}n_{i\downarrow},
\nonumber
\label{singleH}
\end{equation}
where $c^\dagger_{i\sigma}$ is the creation operator of an electron with spin $\sigma$ at site $i$, number operator $n_{i\sigma}=c^\dagger_{i\sigma}c_{i\sigma}$, $i+\delta$ ($i+\delta'$) represents the four first (second) nearest-neighbor sites around site $i$, and $t_h$, $t_h^\prime$, and $U$ are the nearest-neighbor hopping, the next-nearest-neighbor hopping, and on-site Coulomb interaction, respectively.  
We take $U/t_h=10$ and $t_h^\prime/t_h=-0.25$ without being otherwise specified, which are typical values for cuprates with $t_h\sim 0.35$~eV. 

We use a square-lattice periodic Hubbard cluster with $4\times 4$ sites.  
We incorporate an external electric filed via the Peierls substitution in the hopping terms, $c^\dagger_{i,\sigma} c_{j,\sigma} \rightarrow e^{-i\mathbf{A}(t)\cdot\mathbf{R}_{ij}} c^\dagger_{i,\sigma} c_{j,\sigma}$, leading to time-dependent Hamiltonian $H(t)$.  
Here, $\mathbf{A}(t)$ is the vector potential given by
$\mathbf{A}(t)=\mathbf{A}_0 e^{-(t-t_0)^2/(2t_d^2)} \cos[\omega_p(t-t_0)]$, 
where a Gaussian-like envelope centered at $t_0$ has a temporal width $t_d$ and a central frequency  $\omega_p$.  
We apply the external electric field along the $x$ direction, i.e., $\mathbf{A}_0=(A_0,0,0)$ and set $A_0=0.5$, $t_0=0$, $t_d=0.5$, and $\omega_p=U$ without being otherwise specified.  
Hereafter we use $t_h=1$ as an energy unit and $1/t_h$ as a time unit. 

In order to calculate the wave function's time-evaluation $\left|\psi\left( t +\delta t\right)\right\rangle=e^{-iH\left(t\right)\delta t}\left|\psi\left(t\right)\right\rangle$, we employ sequential operations of $H\left(t\right)$~\cite{Shirakawa2020} based on the Taylor expansion:
$\left|\psi\left(t+\delta t\right)\right\rangle\simeq\sum^M_{l=0} \left|\phi_l\right\rangle$,
where $\left|\phi_0\right\rangle=\left|\psi\left(t\right)\right\rangle$ and  $\left|\phi_{l+1}\right\rangle=-iH\left(t\right)\delta t/(l+1)\left|\phi_l\right\rangle$.  
$\delta t$ is set to $0.01$ and $M$ is determined so as to satisfy $\langle\phi_M|\phi_M\rangle<10^{-14}$.

\begin{figure}[t]
\includegraphics[width=0.45\textwidth]{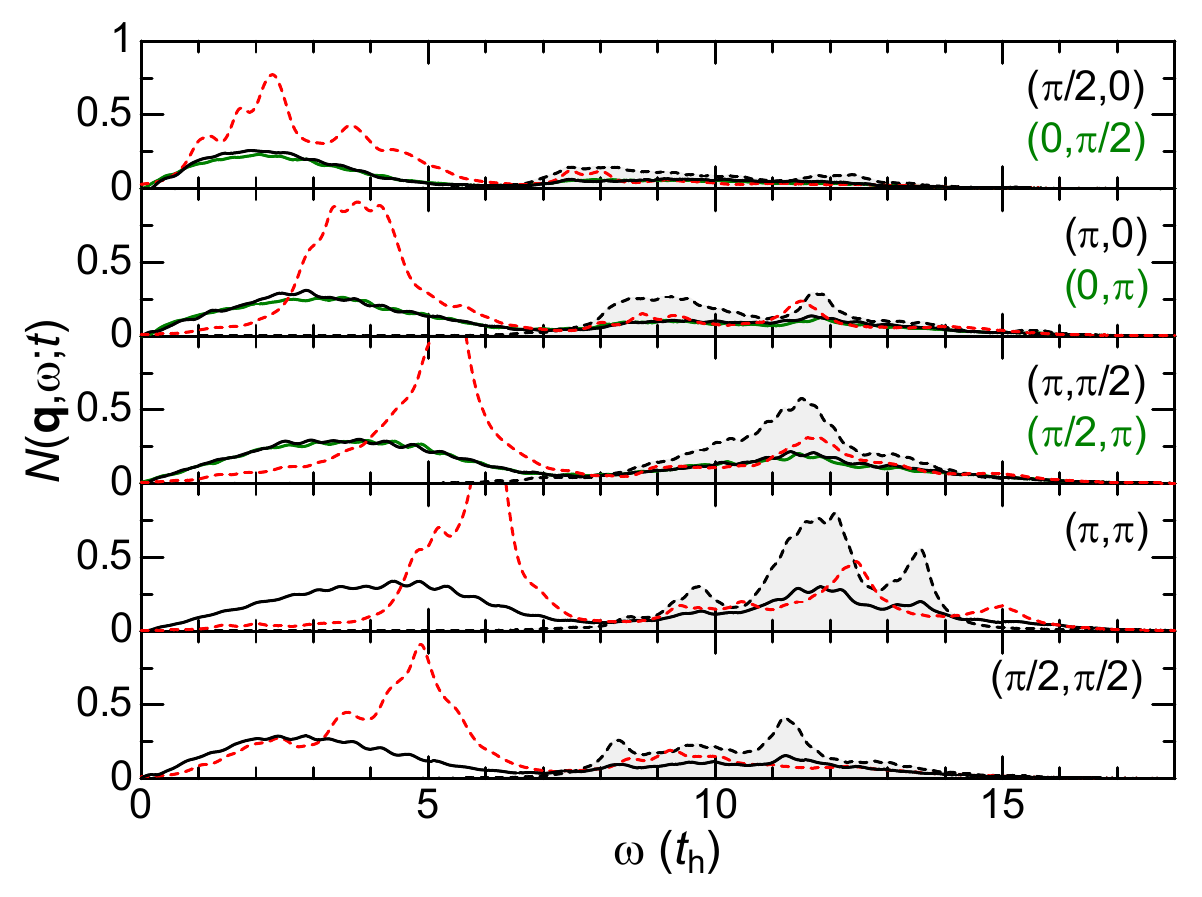}
\caption{Dynamical charge structure factor before and after pumping for the half-filled $4\times 4$ Hubbard lattice with $t_h=1$, $t'_h=-0.25$, and $U=10$.  
The black broken line on each panel represents the equilibrium $N(\mathbf{q},\omega)$ for $\mathbf{q}$ denoted in the panel.  
The black and green solid lines represent $N(\mathbf{q},\omega;t=4)$ for $\mathbf{q}$ denoted by the same color in each panel.  
The red broken line on each panel represents $N(\mathbf{q},\omega)$ for two-hole doped case for the same lattice.}
\label{fig1}
\end{figure}

In calculating the dynamical charge and spin correlation functions in the one-dimensional Hubbard model~\cite{Wang2017} and the $K$-edge RIXS spectrum in the two-dimensional Hubbard model~\cite{Wang2020}, the real-time representation of time-dependent cross section has been used.  
Instead, here we use the spectral representation of dynamical correlation functions regarding $\left|\psi\left(t\right)\right\rangle$ as an initial state, which is easily formulated for the systems without time-dependent terms in their Hamiltonian~\cite{formdcf}.  
Applying this formalism to a time region after turning off the pump pulse, $t>t_\mathrm{off}$, where the Hamiltonian is time-independent, we obtain the time-resolved dynamical correlation function with momentum $\mathbf{q}$ and frequency $\omega$ for a physical quantity $O$ ~\cite{Shinjo2017} as
\begin{eqnarray}
O\left({\bf q},\omega;t\right)
&=&{\rm Im}\sum_{m,n}\frac{1}{\omega+\varepsilon_m-\varepsilon_n+i\eta}
\nonumber\\
&\times&\left[
\left<\psi\left(t\right)\right|\hat{O}_{\bf -q}\left|m\right>
\left<m\right|\hat{O}_{\bf q}\left|n\right>
\left<n|\psi\left(t\right)\right>
\right.\nonumber\\
&-&\left.
\left<\psi\left(t\right)|m\right>
\left<m\right|\hat{O}_{\bf q}\left|n\right>
\left<n\right|\hat{O}_{\bf -q}\left|\psi\left(t\right)\right>
\right],
\label{TrO}
\end{eqnarray}
where $H_0\left| m\right>=\varepsilon_m\left| m\right>$, $\eta$ is a small positive number, and $\left|\psi\left(t\right)\right>=e^{-iH_0(t-t_\mathrm{off})}\left|\psi\left(t_\mathrm{off}\right)\right>$.  
We note that replacing $\left|\psi\left(t\right)\right>$ by the ground state $\left|0\right>=\left|\psi\left(-\infty\right)\right>$ in Eq.~(\ref{TrO}) formally gives the equilibrium dynamical correlation function $O(\mathbf{q},\omega)$.  
We choose $\hat{O}_\mathbf{q}=S^z_\mathbf{q}=\sum_i e^{-i\mathbf{q}\cdot\mathbf{R}_i} S^z_i$ for the time-resolved dynamical spin structure factor $S(\mathbf{q},\omega;t)$ and $\hat{O}_\mathbf{q}=N_\mathbf{q}=\sum_i e^{-i\mathbf{q}\cdot\mathbf{R}_i} N_i$ for the time-resolved dynamical charge structure factor $N(\mathbf{q},\omega;t)$, where $S^z_i$ is the $z$ component of a spin operator and $N_i$ is an electron-number operator at site $i$.  
The integration of the second term in Eq.~(\ref{TrO}) with respect to $\omega$ ($-\infty\leq\omega\leq\infty$) for $S(\mathbf{q},\omega;t)$ [$N(\mathbf{q},\omega;t)$] gives the time-resolved static spin [charge] structure factor $S(\mathbf{q};t)\equiv\left<\psi\left(t\right)\right|S^z_\mathbf{q} S^z_{-\mathbf{q}}\left|\psi\left(t\right)\right>$ [$N(\mathbf{q};t)\equiv\left<\psi\left(t\right)\right|N_\mathbf{q}N_{-\mathbf{q}}\left|\psi\left(t\right)\right>$].


We first discuss charge dynamics that creates the collapse of the Mott gap and forms an in-gap state after photoexcitation.  
Figure~\ref{fig1} demonstrates a change of charge dynamics from the equilibrium $N(\mathbf{q},\omega)$ to $N(\mathbf{q},\omega;t=4)$.  
$N(\mathbf{q},\omega)$ shows momentum-dependent excitations across the Mott gap (colored in gray for each $\mathbf{q}$), which have been observed by Cu $K$-edge RIXS in insulating cuprates~\cite{Hasan2000}.  
After pumping, an in-gap excitation emerges inside the Mott gap below $\omega\sim 7$ with broad spectral weight~\cite{Wang2020}.  
This photoinduced in-gap charge dynamics has been discussed for the same $4\times 4$ Hubbard lattice~\cite{Shinjo2017}.  
The center of gravity for the in-gap spectral weight is momentum-dependent:  
it is lower in energy for small $\mathbf{q}$ and the highest at $\mathbf{q}=(\pi,\pi)$.  
This momentum dependence is qualitatively similar to $N(\mathbf{q},\omega)$ for the two-hole doped $4\times 4$ Hubbard model, but its center of gravity is clearly higher in energy and its spectral distribution is sharper.  
These differences demonstrate a contrasting behavior between chemical-doping and photo-doping.  
Our calculated data indicate that photo-doping induces strong incoherent charge excitation with renormalized energy and broader spectral weight than chemical-doping.  
This may be caused by a drastic change of the electronic state due to photoirradiation.


\begin{figure}[t]
\includegraphics[width=0.45\textwidth]{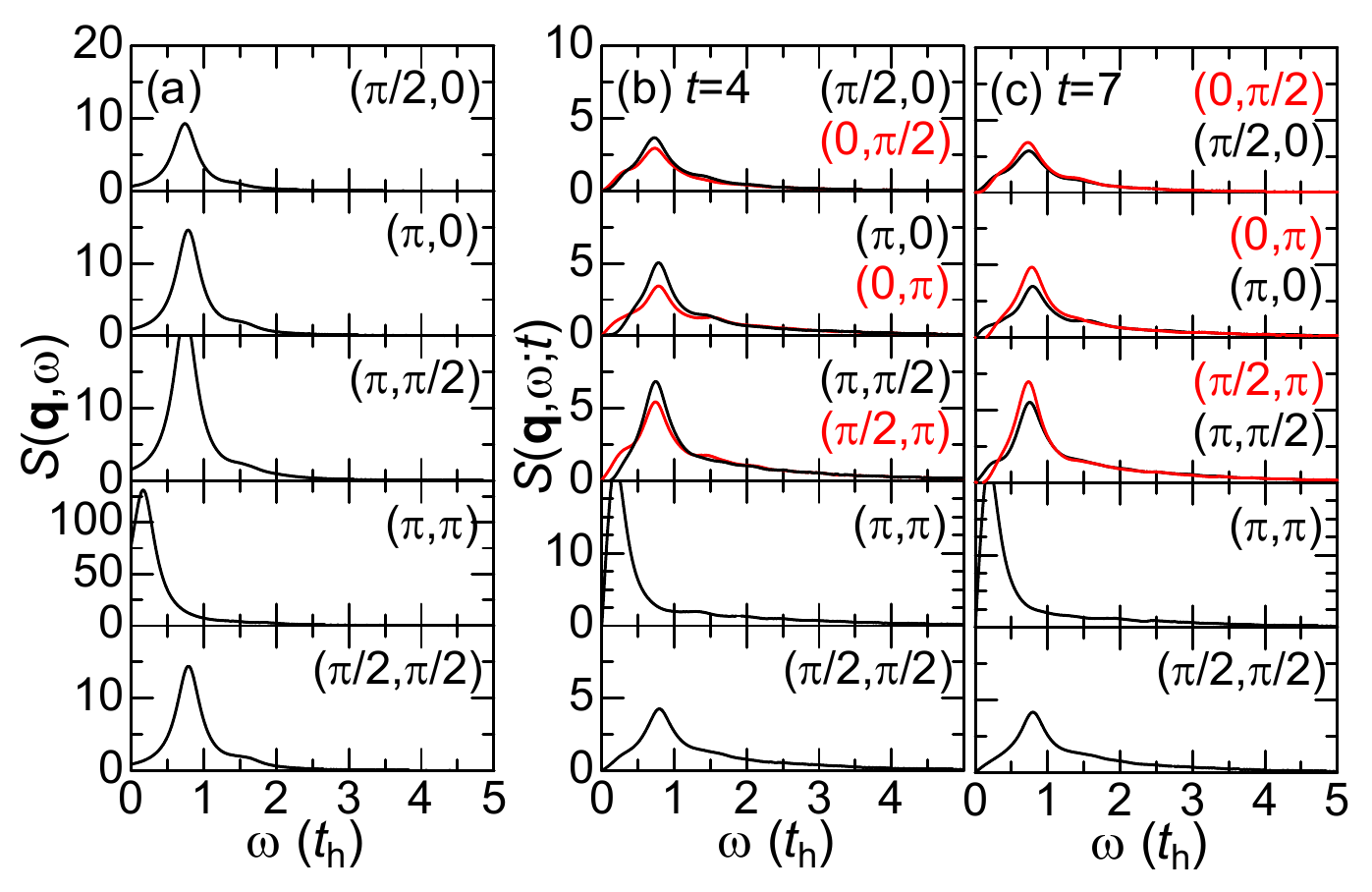}
\caption{Dynamical spin structure factor before and after pumping for the half-filled $4\times 4$ Hubbard lattice with $t_h=1$, $t'_h=-0.25$, and $U=10$.  
(a) $S(\mathbf{q},\omega)$ before pumping for $\mathbf{q}$ denoted in each panel.  
(b) $S(\mathbf{q},\omega;t=4)$.  
The red lines in three panels from the top represent the spectrum for the $\mathbf{q}$ value denoted by red color.  
(c) The same as (b) but for $t=7$.}
\label{fig2}
\end{figure}

In contrast to charge dynamics, spin dynamics does not show a dramatic change in the spectral distribution.  
Figure~\ref{fig2}(a) shows the equilibrium $S(\mathbf{q},\omega)$, where the $\mathbf{q}=(\pi,\pi)$ excitation has the largest intensity.  
After turning off a pump pulse, we find that the spectral intensity decreases but excitation energy is unchanged from $S(\mathbf{q},\omega)$ as shown in Figs.~\ref{fig2}(b) and \ref{fig2}(c) for $S(\mathbf{q},\omega;t=4)$ and $S(\mathbf{q},\omega;t=7)$, respectively.  
However, one can find a notable change of spectral behavior after pumping:  
equivalent momentum points due to reflection symmetry along the $(1,1)$ axis exhibit different weights.  
For example, the spectral intensity at $\mathbf{q}=(\pi,0)$ is larger than that at $\mathbf{q}=(0,\pi)$ in Fig.~\ref{fig2}(b).  
This means that the reflection symmetry is broken.  
Since an electric field is applied along the $x$ direction, this symmetry breaking would be a natural consequence as expected.  
A more interesting observation is that the intensity is strongly time-dependent:  
at $t=7$ the intensity at $\mathbf{q}=(\pi,0)$ is smaller than that at $\mathbf{q}=(0,\pi)$ as seen in Fig.~\ref{fig2}(c).  
This indicates the presence of an oscillating behavior in spectral intensity as a function of $t$. 

\begin{figure}[t]
\includegraphics[width=0.45\textwidth]{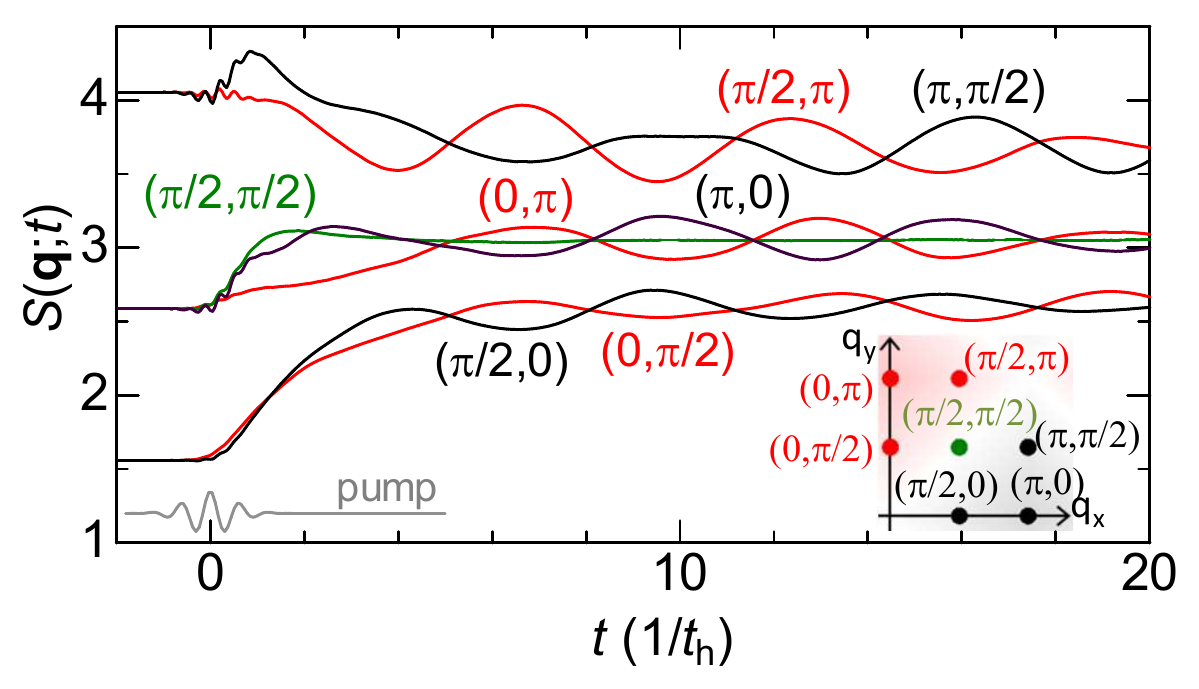}
\caption{Time-resolved static spin structure factor $S(\mathbf{q};t)$ for the half-filled $4\times 4$ Hubbard lattice with $t_h=1$ and $U=10$.  
The color of solid lines corresponds to the color of the $\mathbf{q}$ value denoted in the panel.  
The $\mathbf{q}$ positions are denoted in the inset at the right bottom.  
The profile of the pump pulse is shown at the left bottom, whose amplitude is of arbitrary unit.
}
\label{fig3}
\end{figure}

Since the oscillation of spectral weight causes a change of integrated intensity with respect to $\omega$, it is useful to investigate the time-resolved static spin structure factor $S(\mathbf{q};t)$, which is shown in Fig.~\ref{fig3} for the $4\times 4$ Hubbard lattice with $U=10$ but without $t'_h$.  
Note that $t'_h$ does not change oscillating behaviors in Fig.~\ref{fig3}.  
In calculating $S(\mathbf{q};t)$, we introduce various boundary conditions and average them in order to minimize finite-size effects~\cite{18sites}. 

In Fig.~\ref{fig3}, we find that applying a pumping pulse enhances the magnitude of $S(\mathbf{q};t)$ for small $\mathbf{q}$ [$\mathbf{q}=(\pi/2,0)$, $(0,\pi/2)$, $(\pi,0)$, $(0,\pi)$, and $(\pi/2,\pi/2)$] while it reduces its magnitude for $\mathbf{q}=(\pi,\pi/2)$ and $(\pi/2,\pi)$ as well as for $(\pi,\pi)$ ($\sim40\%$ reduction but not shown here).  
This is because the resonant photoirradiation on the AFM Mott insulator destroys short-ranged AFM correlation, leading to the suppression of spin correlation for large $\mathbf{q}$ values but the enhancement for small $\mathbf{q}$ ones.  
The $S(\mathbf{q};t)$ exhibits oscillations except for $\mathbf{q}=(\pi/2,\pi/2)$.  
The oscillations are antiphase between two momenta equivalent under the reflection with respect to the (1,1) axis.  

\begin{figure}[t]
\includegraphics[width=0.45\textwidth]{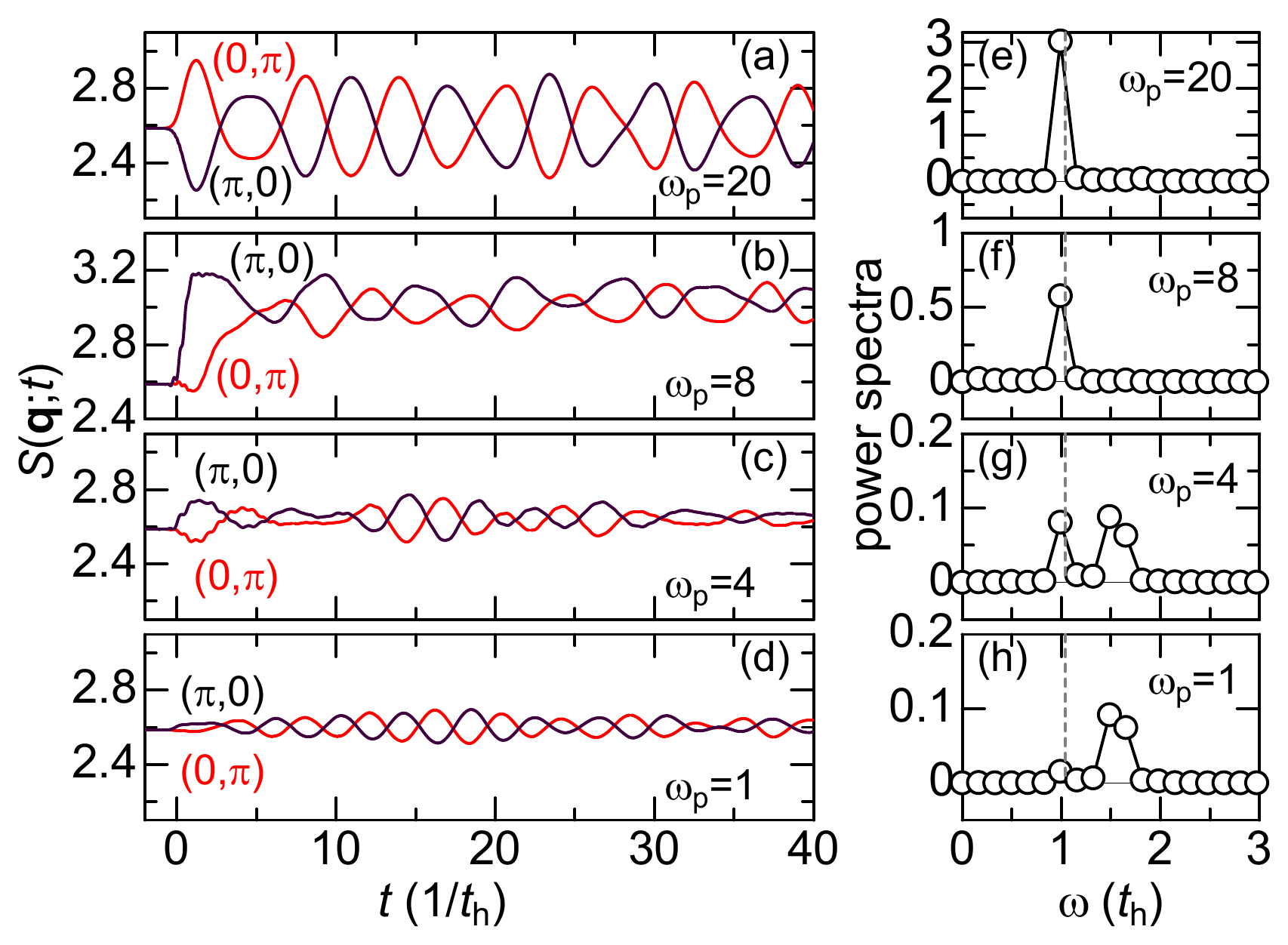}
\caption{Time-resolved static spin structure factor $S(\mathbf{q};t)$ at $\mathbf{q}=(\pi,0)$ and $(0,\pi)$ for the half-filled $4\times 4$ Hubbard lattice with $t_h=1$ and $U=10$.  
(a) $\omega_p=20$, (b) $\omega_p=8$, (c) $\omega_p=4$, and (d) $\omega_p=1$.  
(e) The power spectrum for $\omega_p=20$, (f) for $\omega_p=8$, (g) for $\omega_p=4$, and (h) for $\omega_p=1$.  
The vertical broken line in (e)-(h) represents an eigenenergy with $B_{1g}$ symmetry, where two-magnon Raman intensity is maximized under periodic boundary conditions.}
\label{fig4}
\end{figure}

To clarify the origin of the antiphase oscillations, we focus on two momenta $\mathbf{q}=(\pi,0)$ and $(0,\pi)$ and plot $S(\mathbf{q};t)$ for various $\omega_\mathrm{p}$ in Fig.~\ref{fig4}~\cite{antioscnqw}.  
At $\omega_p=20$ $(>U=10)$, the mean value of $S(\mathbf{q};t)$ with respect to $t$ is almost unchanged from the equilibrium $S(\mathbf{q})$ as shown in Fig.~\ref{fig4}(a), being consistent with an off-resonance condition.  
During pumping centered at $t=0$ (see Fig.~\ref{fig3}),  $S(\mathbf{q}=(\pi,0);t)$ decreases while $S(\mathbf{q}=(0,\pi);t)$ increases.  
The decrease of $S(\mathbf{q}=(\pi,0);t)$ may be understood if one regards the pump pulse as an approximate periodic pulse and applies the Floquet theory with the condition $\omega_p\gg U$, where the leading term of effective AF exchange interaction in the $x$ direction parallel to an applied electric field is given by $J_\parallel^\mathrm{eff}=4t_h^2J_0(A_0)^2/U$ using the zero-th order Bessel function $J_0$ of the first kind~\cite{Mentink2015}.  
$J_\parallel^\mathrm{eff}$ is smaller than the original exchange interaction $J=4t_h^2/U$, leading to the reduction of $S(\mathbf{q}=(\pi,0);t)$ during pumping.  
Contrary, the exchange interaction perpendicular to the electric field, $J_\perp^\mathrm{eff}$, is unchanged from $J$ during pumping.  
Although there is no change of $J_\perp^\mathrm{eff}$, $S(\mathbf{q}=(0,\pi);t)$ increases.  
This increase will be explained by the sum rule of the static spin structure factor that forces to compensate the reduction of $S(\mathbf{q}=(\pi,0);t)$ by enhancing $S(\mathbf{q}=(0,\pi);t)$. 

At near-resonance $\omega_\mathrm{p}\lesssim U$ [Fig.~\ref{fig4}(b), also see Fig.~\ref{fig3}], $S(\mathbf{q}=(\pi,0);t)$ increases quicker than $S(\mathbf{q}=(0,\pi);t)$ during pumping.  
This is an opposite behavior to the case of $\omega_p=20>U$.  
It is crucial to notice that resonant driving by a periodic electric field with $\omega_p=U$ gives rise to an additional process due to doublon association and dissociation~\cite{Bukov2016}.  
The additional process induces a new contribution to the exchange interaction given by $4t_h^2J_1(A_0)^2\{1/(U-\omega_p)+1/(U+\omega_p)\}$~\cite{Desbuquois2017,Gorg2018} for $\omega_p\lesssim U$ on top of nonresonant contribution $4t_h^2J_0(A_0)^2/U$, where $J_1$ is the first-order Bessel function of the first kind.  
This new contribution enhances $J_\parallel^\mathrm{eff}$, resulting in the increase of $S(\mathbf{q}=(\pi,0);t)$ during pumping.
 
The discussion based on the Floquet theory can also be applied to another off-resonance case with $t_h\ll\omega_p\ll U$, which corresponds to the case of $\omega_p=4$ in Fig.~\ref{fig4}(c).  
In this case, $S(\mathbf{q}=(\pi,0);t)$ increases during pumping.  
This increase can be understood by the enhancement of $J_\parallel^\mathrm{eff}$ due to the Floquet side band in the weak excitation regime~\cite{Mentink2015}.  
$S(\mathbf{q}=(0,\pi);t)$ decreases due to the sum rule.  
We find a qualitatively similar change during pumping even for the case of $\omega_p=1$ as shown in Fig.~\ref{fig4}(d). 

Now let us discuss the antiphase oscillations at $\mathbf{q}=(0,\pi)$ and $(0,\pi)$ after turning off a pump pulse.  
At $\omega_p=20$, we find oscillations with period $\sim 6$ [see Fig.~\ref{fig4}(a)].  
This period produces a peak at $\omega=1$ in the power spectrum as shown in Fig.~\ref{fig4}(e).  
We find that the frequency of $\omega=1$ agrees with an eigenenergy in the $B_{1g}$-symmetry subspace with zero total momentum of $H_0$, which is denoted by the vertical broken line in Figs.~\ref{fig4}(e)-\ref{fig4}(h).  
This means that the oscillation is Rabi-type between the ground state and the $B_{1g}$ eigenstate.   The $B_{1g}$ eigenstate is a Raman active state in $H_0$ and gives the highest spectral weight in a two-magnon Raman process~\cite{Tohyama2002}.  
Therefore, it is reasonable to consider that photoexcited states after pumping may predominantly consist of the bases with $B_{1g}$ symmetry.  
This reasoning naturally leads to a contribution of the $B_{1g}$ eigenstate $\left|B_{1g}\right>$ with energy $\varepsilon_{B_{1g}}$ to $S(\mathbf{q};t)=\sum_{mn}e^{i(\varepsilon_m-\varepsilon_n)t}\left<\psi(t_\mathrm{off})\right|m\left>\right<m\left|S^z_\mathbf{q}S^z_{-\mathbf{q}}\right|n\left>\right<n\left|\psi(t_\mathrm{off})\right>$, which is given by
\begin{eqnarray}
S_{B_{1g}}(\mathbf{q};t)&=& e^{i(\varepsilon_{B_{1g}}-\varepsilon_0)t}\left< 0\right|\psi(t_\mathrm{off})\left>\right<\psi(t_\mathrm{off})\left| B_{1g}\right> \nonumber \\
&&\times\left<B_{1g}\right|S^z_\mathbf{q}S^z_{-\mathbf{q}}\left| 0\right> + \mathrm{c.c.}. \nonumber
\end{eqnarray}
This gives rise to an oscillation with frequency $\varepsilon_{B_{1g}}-\varepsilon_0$.  
In addition, the reflection along the (1,1) axis $\hat{\sigma}_d$ applied to $\mathbf{q}$ yields $\left<B_{1g}\right|S^z_{\hat{\sigma}_d(\mathbf{q})}S^z_{\hat{\sigma}_d({-\mathbf{q}})}\left| 0\right>=-\left<B_{1g}\right|S^z_\mathbf{q}S^z_{-\mathbf{q}}\left| 0\right>$, leading to $S_{B_{1g}}(\hat{\sigma}_d(\mathbf{q});t)=-S_{B_{1g}}(\mathbf{q};t)$.  
This is the origin of antiphase oscillations between $S(\mathbf{q}=(\pi,0);t)$ and $S(\mathbf{q}=(0,\pi);t)$.  
In other words, a dominant contribution of the $B_{1g}$ excited states to the time-dependent wave function leads to out-of-phase oscillations in the spin structure factor at the originally equivalent two momentum positions, $\mathbf{q}=(\pi,0)$ and $(0,\pi)$.  
This phase difference is consistent with the sign difference of an $x^2-y^2$ function with the $B_{1g}$ symmetry between $(x,y)=(\alpha,0)$ and $(0,\alpha)$.

A remaining question is why the $B_{1g}$ eigenstate is predominately selected during pumping.  
We should emphasize that the $B_{1g}$ eigenstate is the final state of a two-magnon Raman process, where magnetic excitation is minimized in energy among other excited states.  
This means that off-resonant photoexcitation selects a magnetically stable state.

With decreasing $\omega_p$, the power spectrum has a peak at $\omega=1$ as shown in Figs.~\ref{fig4}(f) and \ref{fig4}(g).  
In addition, there is another peak at $\omega=1.5$ in Figs.~\ref{fig4}(g) and \ref{fig4}(h).  
There are both the $B_{1g}$ and $A_{1g}$ eigenstates near $\omega=1.5$.  
Therefore, it is not easy to assign which eigenstate contributes to the peak position, but the $B_{1g}$ eigenstate will be more probable because of the presence of antiphase oscillations even for $\omega_\mathrm{p}=1$~\cite{inphasea1g}.  

The antiphase oscillations in spin correlation between $\mathbf{q}=(q, 0)$ and $(0,q)$ seem to be a characteristic effect in photoexcitaed AFM Mott insulator.  
It would be possible to detect such oscillations in $S(\mathbf{q},\omega;t)$ by the $L$-edge TRRIXS~\cite{REXS} for cuprate and iridate AFM Mott insulators if the present experimental time resolution ($\sim 400$fs~\cite{Mazzone2020}) is improved by more than an order of magnitude.  
As discussed above, the antiphase oscillations are clear evidence for the presence of photoexcited $B_{1g}$ states if an electric field is applied along the direction of AFM exchange interaction in a square lattice.  
Therefore, TRRIXS that can detect this evidence will develop as one of novel techniques to clarify the symmetry of photoexcited states.

In summary, we have numerically examined momentum-dependent spin and charge excitations after turning off a pumping pulse in the AFM Mott insulator on a square lattice described by the half-filled Hubbard model.  
We have clearly found novel momentum-dependent transient spin dynamics, where the temporal oscillations of the intensity of spin excitation show an antiphase behavior for two orthogonal directions that are parallel and perpendicular to the electric field of a pump pulse.  
Their oscillation period in time is determined by the excitation energy of two magnons in the AFM Mott insulator and the phase difference comes from the $B_{1g}$ channel of the two-magnon excitation.  
This result is a theoretical prediction and thus observing this prediction will be a challenging subject for TRRIXS experiment when its time resolution is improved by more than an order of magnitude.


This work was supported by QST President's Strategic Grant(QST Advanced Study Laboratory), by the Japan Society for the Promotion of Science, KAKENHI (Grants No. 19H01829 and No. JP19H05825) from Ministry of Education, Culture, Sports, Science, and Technology, Japan, and by CREST (Grant No. JPMJCR1661), the Japan Science and Technology Agency, Japan. Part of the computational work was performed using the supercomputing facilities in JAEA.


\nocite{*}


\begin{thebibliography}{99}
\bibitem{formdcf}This representation contains the double sum of all eigenstates instead of the single sum as in the standard spectral representation, because the initial state $\left|\psi\left(t\right)\right\rangle$ is not a pure eigenstate.
\bibitem{18sites}We also calculated $S(\mathbf{q};t)$ in an 18-site Hubbard cluster with several boundary conditions and found oscillating behaviors similar to the 4$\times$4 cluster.
\bibitem{antioscnqw}We also found the antiphase oscillations in $N(\mathbf{q};t)$.  
However, at $\omega_\mathrm{p}=20$ the oscillation amplitude relative to $N(\mathbf{q}=(\pi,0);t)$ was found to be 3\% and thus smaller than 9\% for $S(\mathbf{q};t)$ as seen in Fig.~\ref{fig4}(a).  
At $\omega_\mathrm{p}=10$, the amplitude is only 1\%, being consistent with very small difference in $N(\mathbf{q},\omega;t)$ between $\mathbf{q}=(\pi,0)$ and $(0,\pi)$ as seen in Fig.~\ref{fig1}.  
We also found that, when $\omega_\mathrm{p}$ is less than the value of the Mott gap, such oscillations are hardly seen and another oscillation with a shorter period coupled to the Mott-gap excitation emerges.
\bibitem{inphasea1g}If the oscillation comes from the $A_{1g}$ eigenstate, there should be an in-phase behavior because $\left<A_{1g}\right|S^z_{\hat{\sigma}_d(\mathbf{q})}S^z_{\hat{\sigma}_d({-\mathbf{q}})}\left| 0\right>=\left<A_{1g}\right|S^z_\mathbf{q}S^z_{-\mathbf{q}}\left| 0\right>$.
\end{thebibliography}

\end{document}